\documentclass[a4paper]{article}

\usepackage{INTERSPEECH2022}
\usepackage{CJKutf8}
\usepackage{graphicx}
\usepackage{subcaption}
\usepackage{hyperref}
\usepackage{amssymb}
\usepackage{multirow}
\usepackage{booktabs}
\usepackage{url}

\title{A Study of Modeling Rising Intonation in Cantonese Neural Speech Synthesis}
\name{Qibing Bai$^1$, Tom Ko$^{2,*}$\thanks{* corresponding author}, Yu Zhang$^1$}
\address{
  $^1$Department of Computer Science and Engineering, Southern University of Science and Technology \\
  $^2$ByteDance AI Lab
}
\email{}

\begin{document}

\maketitle
\ninept
\begin{abstract}
In human speech, the attitude of a speaker cannot be fully expressed only by the textual content. It has to come along with the intonation.
Declarative questions are commonly used in daily Cantonese conversations, and they are usually uttered with rising intonation.
Vanilla neural text-to-speech (TTS) systems are not capable of synthesizing rising intonation for these sentences due to the loss of semantic information.
Though it has become more common to complement the systems with extra language models, their performance in modeling rising intonation is not well studied.
In this paper, we propose to complement the Cantonese TTS model with a BERT-based statement/question classifier.
We design different training strategies and compare their performance.
We conduct our experiments on a Cantonese corpus named CanTTS.
Empirical results show that the separate training approach obtains the best generalization performance and feasibility.

\end{abstract}


\section{Introduction}
\label{sec:intro}
\vspace{4mm}

Significant developments of neural text-to-speech (TTS) synthesis have achieved realistic speech generation \cite{shen2018natural,ren2019fastspeech,kim2021conditional}.
Neural TTS models are based on deep neural networks and trained with large corpora. Using an encoder-decoder architecture, neural TTS models map input characters or phonemes to acoustic features (e.g., mel-spectrograms) or directly to the waveform. The acoustic features can be converted to waveforms with vocoders \cite{vanwavenet,yang2021multi}.
Though synthesized speech is close to human speech in naturalness, its prosody is not always suitable given the text.

For prosody modeling, traditional parametric synthesis explicitly represents prosody components with models such as ToBI \cite{silverman1992tobi} via text analysis. Neural TTS methods usually learn prosody implicitly, and the generated prosody often sounds flattened.
In order to improve the generated prosody, several variational \cite{hsu2019hierarchical,zhang2019learning} and non-variational \cite{skerry2018towards,wang2018style} methods have been proposed to learn latent prosodic representations.
Some methods \cite{vioni2021prosodic,gong2021improving} are proposed for low-level prosody control.
In inference, though some prosody attributes like emotions can be captured and transferred through reference audios, other prosody attributes related to the context could be inappropriate.
Several works exploit various syntactic and semantic features in neural TTS \cite{guo2019exploiting,song2021syntactic,hayashi2019pre,xiao2020improving} to generate prosody suitable for the context.
A pre-trained BERT-like speech module \cite{chen2021speech} learns prosodic patterns from unlabeled speech data to improve synthesized prosody.
\cite{zou2021fine} leverages ToBI labels into neural TTS to improve the prosody.
Recent works \cite{hodari2021camp,karlapati2021prosodic} attempt to sample from the learned prosodic distribution using contextual information.
However, sentence-level prosody related to the speaker's intention draws much less attention.

A speaker's attitude can not be fully expressed only by phoneme segments in speech. It has to come along with the intonation.
For example, a speaker can change a statement into a declarative question\footnote{In Chinese literature, declarative questions are also referred to as unmarked questions.} with rising intonation.
A declarative question usually corresponds to a yes/no answer.
Consider the following sentences\footnote{English translations may have different intonation.}:
\begin{itemize}
    \item \begin{CJK*}{UTF8}{gbsn}他去学校。\end{CJK*}
    He goes to school.  (A statement)
    \item \begin{CJK*}{UTF8}{gbsn}他去学校？ \end{CJK*}
    He goes to school? (A declarative question)
    \item \begin{CJK*}{UTF8}{gbsn}他去不去学校？ \end{CJK*}
    Does he go to school? (A normal question)
\end{itemize}

Rising intonation is not necessary for a normal question as listeners can understand that it is a question from spoken phonemes.
Therefore, speakers tend to utter normal questions with intonation similar to statements, especially in languages like Mandarin and Cantonese, where question particles play the main role.
Although declarative questions are ended with the question mark in their written form, if they are uttered with a normal intonation, the listeners may not perceive the speaker's intention correctly.
Thus, rising intonation is necessary for intention clarification.

Basic neural TTS models that solely take phonemes as input are not capable of synthesizing rising intonation for declarative questions.
It is because of two reasons.
Firstly, questions constitute only a small portion of available data corpora, not to mention declarative questions. Thus, samples with rising intonation are insufficient for a model to learn implicitly.
Secondly, the grapheme-to-phoneme (g2p) conversion causes a loss of semantic and syntactic information that identifies the sentence type. 
g2p helps generalize on rare/unseen characters and is necessary for languages with a large number of characters, e.g., Chinese and Korean.
Besides, it is hard for TTS models to capture the sentence type without powerful language models.
Thus the TTS model alone cannot map the input text to its associated intonation.

The judgment of declarative questions associated with rising intonation depends on the sentence semantics. However, various methods \cite{hsu2019hierarchical,zhang2019learning,skerry2018towards,wang2018style} based on sampling or reference audios do not consider textual semantic information, nor does the pre-trained prosody encoder \cite{chen2021speech}.
Recent attempts on injecting various linguistic features \cite{guo2019exploiting,song2021syntactic,hayashi2019pre,xiao2020improving} and context-based prosody sampling \cite{hodari2021camp,karlapati2021prosodic} can compensate for the information loss, but they do not provide discriminative information about sentence types. Therefore, the performance on rising intonation featured by declarative questions is not well studied yet.

In this paper, we propose to complement the TTS model with a BERT-based statement/question classifier.
We design three training approaches: two are named ``explicit'' because they use sentence type labels in training, and the other requires no label and is called ``implicit''.
We utilize a BERT-based sentence classifier to provide the sentence type information to the TTS model.
We conduct our experiments on a Cantonese corpus named CanTTS, which explicitly labels the sentence type and intonation.
It contains three types of sentences: statements, normal questions, and declarative questions.
The statements and normal questions are all uttered with non-rising intonation; the declarative questions are all uttered with rising intonation.
Objective and subjective evaluations show that all models augmented with the sentence classifier outperform the baseline Tacotron2. A separately trained classifier achieves the best performance while enjoying the feasibility of leveraging extra text data.


\section{Related Work}
\label{sec:review}
\vspace{4mm}

BERT \cite{devlin-etal-2019-bert} is a language model that learns to capture contextualized text information and generate general-purpose text representations. It is trained on large text corpora with masked language modeling. The generated text representations carry both syntactic and semantic information \cite{jawahar2019does} that is useful in prosody generation.
Though the $\rm[CLS]$ token is supposed to capture sentence-level information, BERT usually needs extra discriminative guidance when applied to classification tasks.

In \cite{hayashi2019pre}, an additional attention module is used to extract context vector for the Tacotron \cite{wang2017tacotron} decoder. \cite{hayashi2019pre} also directly concatenates the $\rm[CLS]$ token with phoneme representations and feeds them to the decoder. \cite{tyagi2020dynamic} uses BERT to provide semantic features in comparison with syntactic features. They average representations from multiple BERT layers, which are used to select a sentence-level acoustic embedding. \cite{xiao2020improving} uses BERT-derived representations to predict breaks as extra inputs to the encoder. They also test directly concatenating the representations to phoneme embeddings. \cite{xu2021improving} uses contextual representations of neighboring sentences for prosody improvement. They use a cross-utterance encoder to process BERT-derived representations and concatenate them with phoneme representations. PnG BERT \cite{jia2021png} is a TTS encoder that takes characters and phonemes as its input. It is pre-trained on text data with phoneme transcriptions, therefore able to provide both pronunciation and semantic information.

All the above approaches are shown to improve the naturalness of synthesized speech. However, they do not provide adequate information about declarative questions, which features rising intonation in Cantonese.

\section{Our Approach}
\label{sec:method}
\vspace{4mm}

\begin{figure}[h]
\begin{center}
\includegraphics[width=0.90\linewidth]{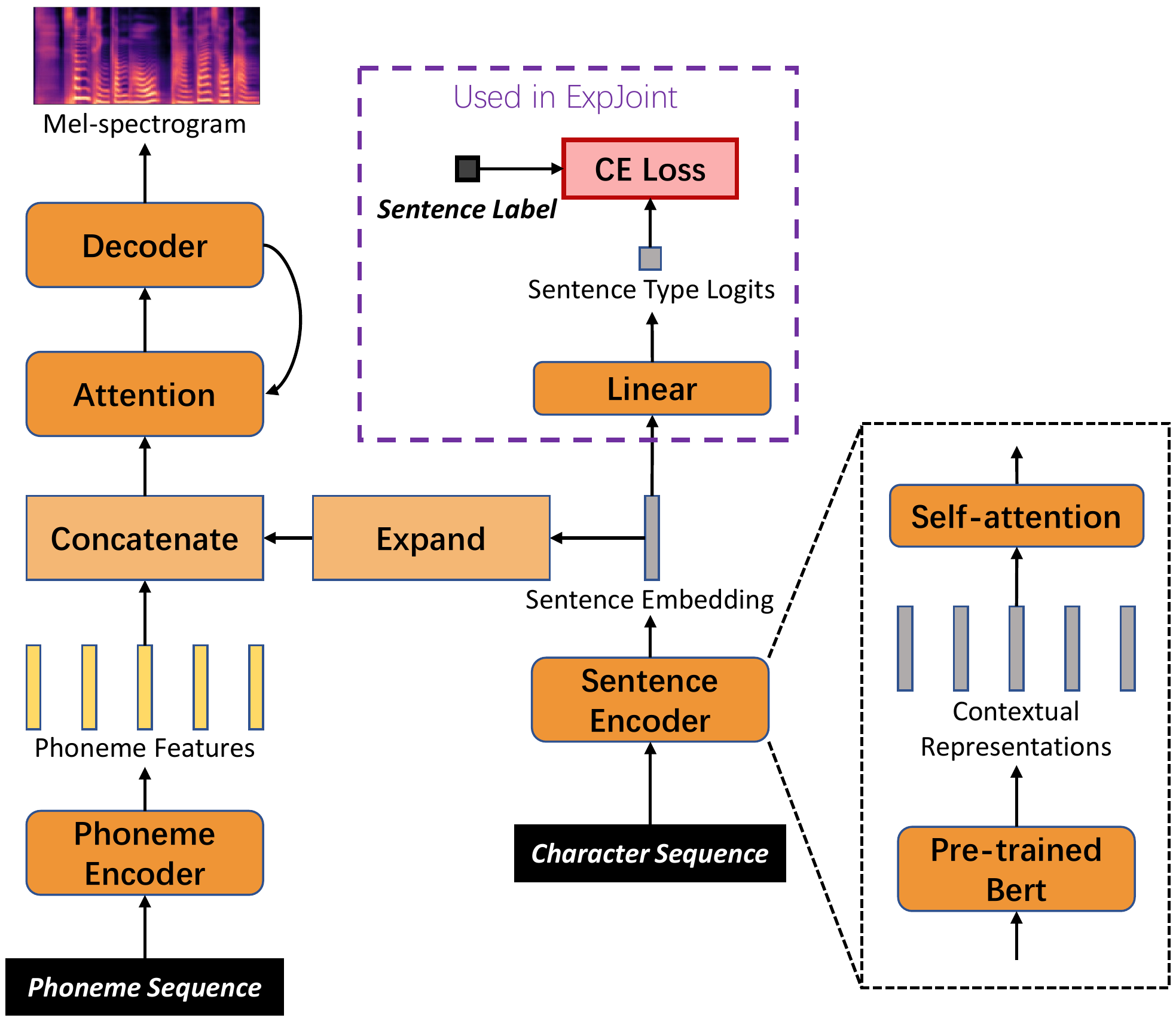}
\end{center}
   \caption{Structure of ImpJoint and ExpJoint. The modules in the black dashed box make up the sentence encoder. The modules in the purple dashed box are used in ExpJoint training.}
\label{fig:exp_joint}
\end{figure}

We design a BERT-based statement/question classifier to provide the sentence type information associated with non-rising/rising intonation.
The sentence classifier consists of a sentence encoder and a classification layer.
All the systems in this work are based on Tacotron2 and the sentence encoder/classifier.
The backbone of Tacotron2 is shown in the left part of Fig. \ref{fig:exp_joint}. We additionally use the MAE loss and the guided attention loss \cite{tachibana2018efficiently} for faster convergence. We also adopt Multi-band MelGAN \cite{yang2021multi} as the vocoder.

We examine an implicit approach and two explicit approaches for learning the intonation from the data.
The implicit approach does not require sentence labels in training, while the explicit ones do.

\subsection{Sentence encoder/classifier}
\label{subsec:sen_enc}

The sentence encoder is shown in the black dashed box in Fig. \ref{fig:exp_joint}. It consists of a pre-trained BERT model and a self-attention module. The pre-trained BERT model, which is powerful in extracting semantic information, outputs token-level representations. 
The self-attention module extracts a sentence-level embedding from a linear transformation of the token-level representations \cite{zhu2018self}.
The formulations of the self-attention module are as follows.
We first compute a scalar score $e_t$ for each input representation ${\textbf h}_t$, normalize the scores to obtain weights $\alpha_t$, and finally sum up the representations:
\begin{align}
    e_t &= {\textbf v}^T {\rm tanh} ({\textbf W} {\textbf h}_t + {\textbf b})\\
    \alpha_t &= \frac{{\rm exp}(e_t)}{\sum_{i=1}^{T} {\rm exp}(e_i)}\\
    {\textbf s} &= \sum_{t=1}^{T} \alpha_t {\textbf h}_t,
\end{align}
where ${\textbf s}$ is the sentence embedding, ${\textbf h}_t$ is the $t$-th output of the self-attention layers, and transformation parameters ${\textbf W}$, ${\textbf b}$ as well as the query vector ${\textbf v}$ are learnable.

For sentence-level tasks like sentence classification, as suggested in \cite{devlin-etal-2019-bert}, it is common to use the classifier embedding $\rm[CLS]$ alone with a classification layer. In our work, we empirically find that a linear combination of the $\rm[CLS]$ and all other contextual representations can lead to better overall results. Thus, we apply this setup in all the systems. In our work, the $\rm[CLS]$ token and the contextual representations are derived from a pre-trained Cantonese BERT.

The sentence encoder can be followed by an additional classification layer for statement/question classification, as shown in the purple dashed box in Fig. \ref{fig:exp_joint}.
There are three sentence categories: statement, normal question, and declarative question.
If we remove end punctuation marks, the problem becomes binary classification (normal question vs. statement/declarative question). We still adopt the three classes since it is trivial for BERT to consider end punctuation marks.

\subsection{Implicit approach -- ImpJoint}

In this approach, it is assumed that there are certain utterances with rising intonation, but no sentence labels are available.
This simulates the real situation where labels on prosody are difficult to acquire. 
We intend to investigate whether the TTS model can implicitly learn contextual information of rising intonation.

The model architecture is shown in Fig. \ref{fig:exp_joint}.
The input text is processed by the sentence encoder to generate a sentence embedding, which is then expanded and concatenated to encoded phoneme features.
The Tacotron2 and the sentence encoder are jointly trained.

\subsection{Explicit approach}

In this approach, it is assumed that we can make use of the sentence and intonation labels in training.
We design and examine two explicit approaches: a joint training approach and a separate training approach.

\subsubsection{Joint training -- ExpJoint}

We utilize the sentence labels in the joint training framework by incorporating the sentence classifier to perform an extra classification task.
The sentence embedding is fed to both the TTS model and the classification layer (shown in the purple dashed box in Fig. \ref{fig:exp_joint}).
The cross-entropy (CE) loss is used.

We expect that the sentence embeddings can carry more discriminative information about the sentence type.
This helps the TTS model to learn the relationship between the sentence type and the intonation in the spectrogram.

\begin{figure}[h]
\centering
\includegraphics[width=0.90\linewidth]{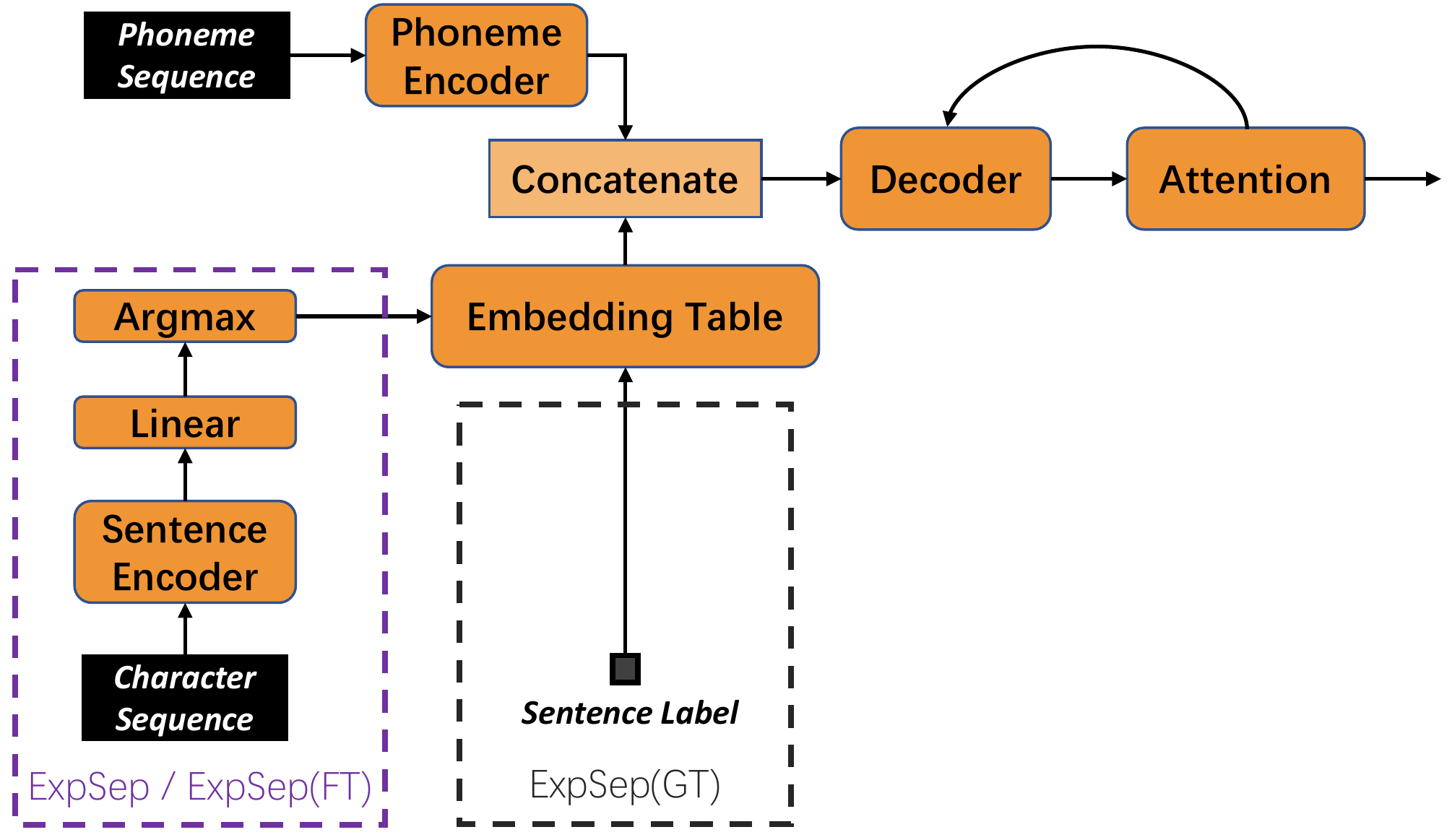}
\caption{Structure of ExpSep, ExpSep(FT) and ExpSep(GT).}
\label{fig:exp_sep}
\end{figure}

\subsubsection{Separate training -- ExpSep}

As the sentence and intonation labels are available, it is feasible to train the TTS model and the sentence classifier separately (i.e., Fig. \ref{fig:exp_sep}).
The TTS model and the sentence classifier are connected with an embedding table where each entry corresponds to a sentence type.

A major advantage of the separate training approach is that the sentence classifier can be trained with extra text data without paired speech. It is much easier to collect text than to prepare paired speech data.
We basically freeze the BERT parameters for comparison with joint training approaches. We also test fine-tuning the BERT parameters while training the classifier and name this model ExpSep(FT).

If we allow the user to hint at the sentence type, the TTS model separately trained in this approach can actually be used alone.
We name this TTS model ExpSep(GT) as we assume that it is always fed with ground-truth sentence labels.

\section{Experimental Setup}
\label{sec:exp}


\subsection{Dataset}

We conduct our experiments on CanTTS\footnote{\url{https://github.com/parami-ai/CanTTS}}, a single-speaker Cantonese corpus consisting of 12,010 utterances.
The corpus contains $\sim$10,000 statements and $\sim$2000 questions.
The question set can be further divided into $\sim$1,000 normal questions and $\sim$1,000 declarative questions.
There is a total of 20 hours of speech recorded from a female speaker aged 20.
All the statements are ended up with either a comma or a full stop, and all the questions are ended up with a question mark in their transcription.
In our work, all the punctuation marks are processed as part of the input text.
All the declarative questions are uttered with rising intonation.
All the statements and normal questions are uttered with a non-rising intonation.

Some of the utterances contain English words, so we exclude these code-mixed samples. We further divide the corpus into a training set and a test set. The proportions of sentence types are the same across the two subsets. The test set contains 448 statements, 50 normal questions, and 50 declarative questions.

\subsection{Model setting}

We use the pre-trained Cantonese BERT-Base from Huggingface \cite{wolf2020transformers} and conduct our experiments on ESPnet \cite{hayashi2020espnet}.
As explained in Section \ref{subsec:sen_enc}, the sentence classifier has 3 output types, and the sentence embedding table in ExpSep, ExpSep(FT) and ExpSep(GT) has 3 entries. The dimension of the embedding table is 512.

All TTS models are trained with the Adam optimizer \cite{kingma2015adam}, a batch size of 135, and a learning rate of $10^{-4}$. In training ExpJoint, the weight of sentence classification loss is 0.1.
The class weights of statements, normal questions, and declarative questions within the classification loss are 1, 10, and 20.
In training the sentence classifier in ExpSep and ExpSep(FT), the learning rate is $10^{-5}$ and the batch size is 512.
We expand the text training set by removing the end punctuation marks of all sentences; the declarative questions without end punctuation are taken as statements.

\section{Results and Discussion}

All designed models, with the help of the sentence encoder/classifier, can synthesize rising intonation for declarative questions.
To further compare the model performance, we use both objective and subjective evaluations. For ease of presentation, statements, normal questions, and declarative questions are abbreviated Sta, Que, and DecQue, respectively.
On the testing set, ExpSep(FT) predicts labels for all the sentences. Therefore, we present performance of the two systems with the name ExpSep(GT/FT).

\begin{figure}[ht]
\centering
\begin{subfigure}{0.22\textwidth}
    \includegraphics[width=\linewidth]{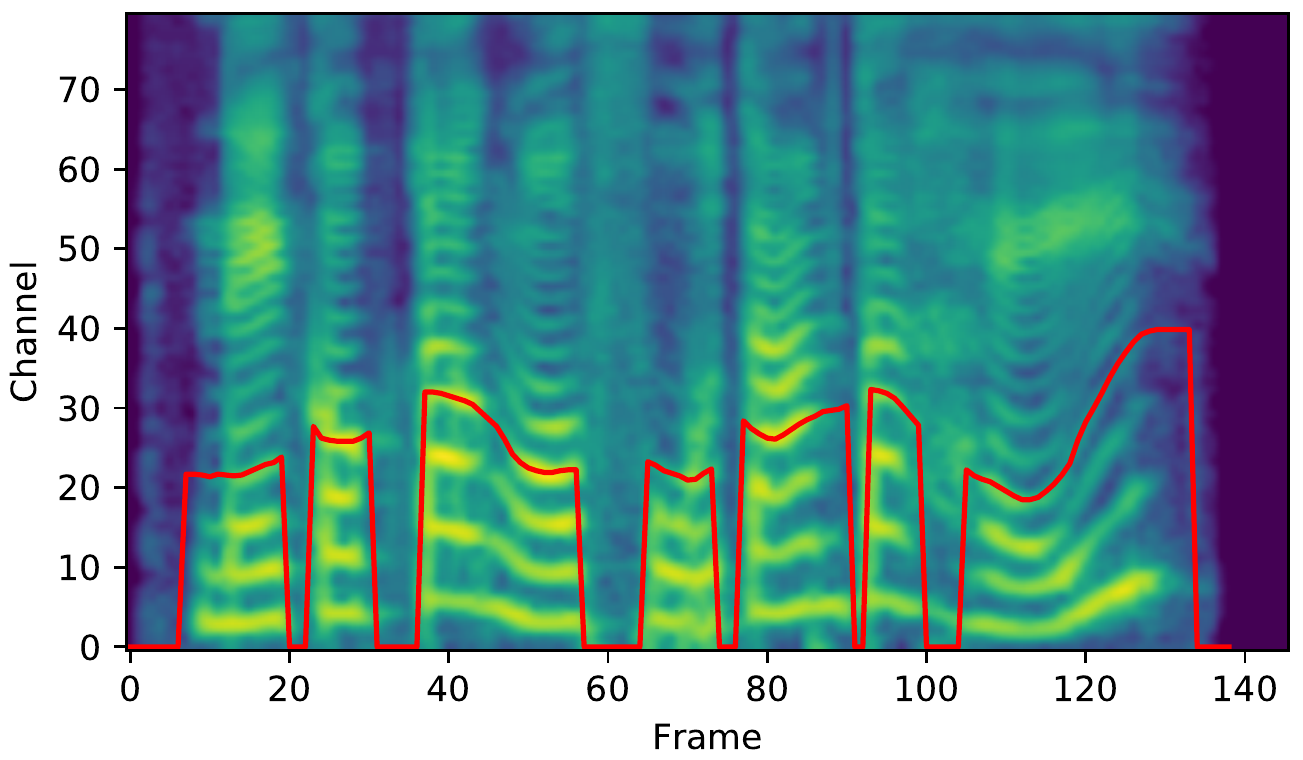}
    \caption{\normalfont Ground truth}
    \label{fig:lmspc_gt}
\end{subfigure}
\begin{subfigure}{0.22\textwidth}
    \includegraphics[width=\linewidth]{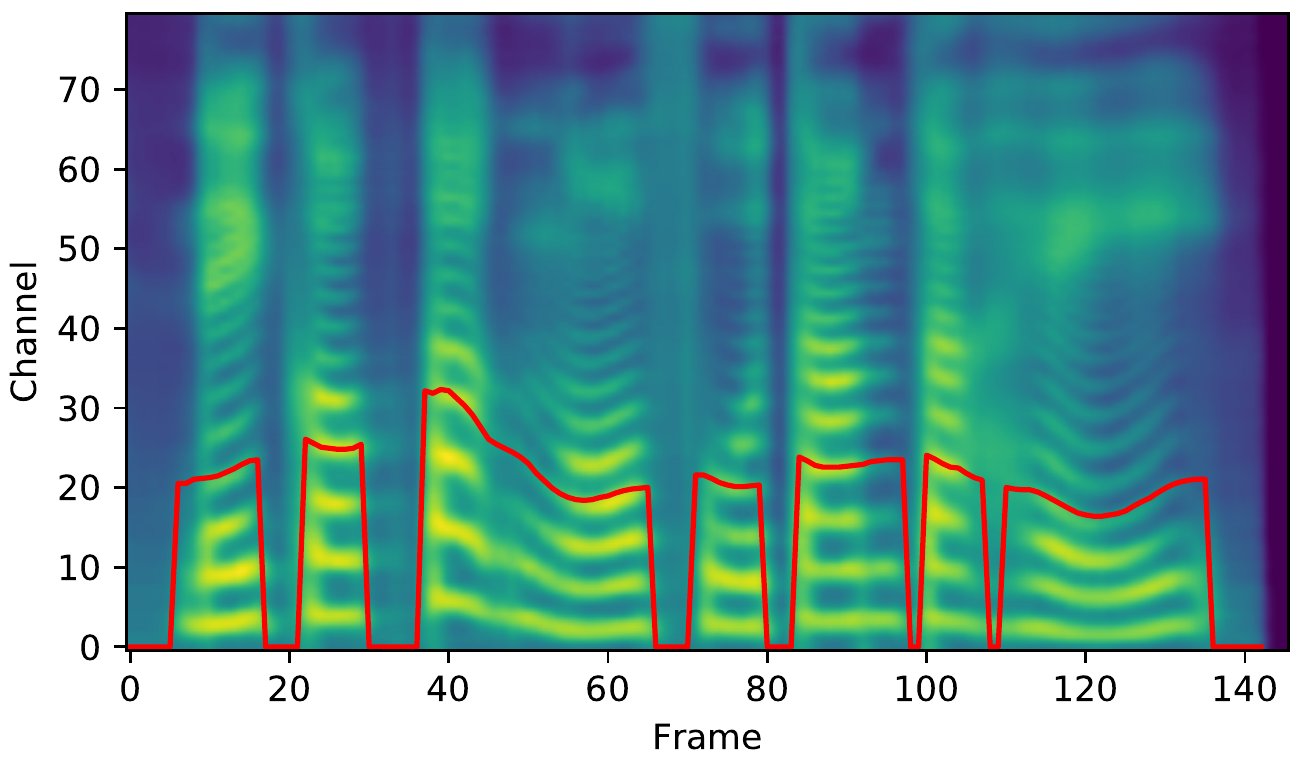}
    \caption{\normalfont Tacotron2}
    \label{fig:lmspc_tacotron2}
\end{subfigure}
\begin{subfigure}{0.22\textwidth}
    \includegraphics[width=\linewidth]{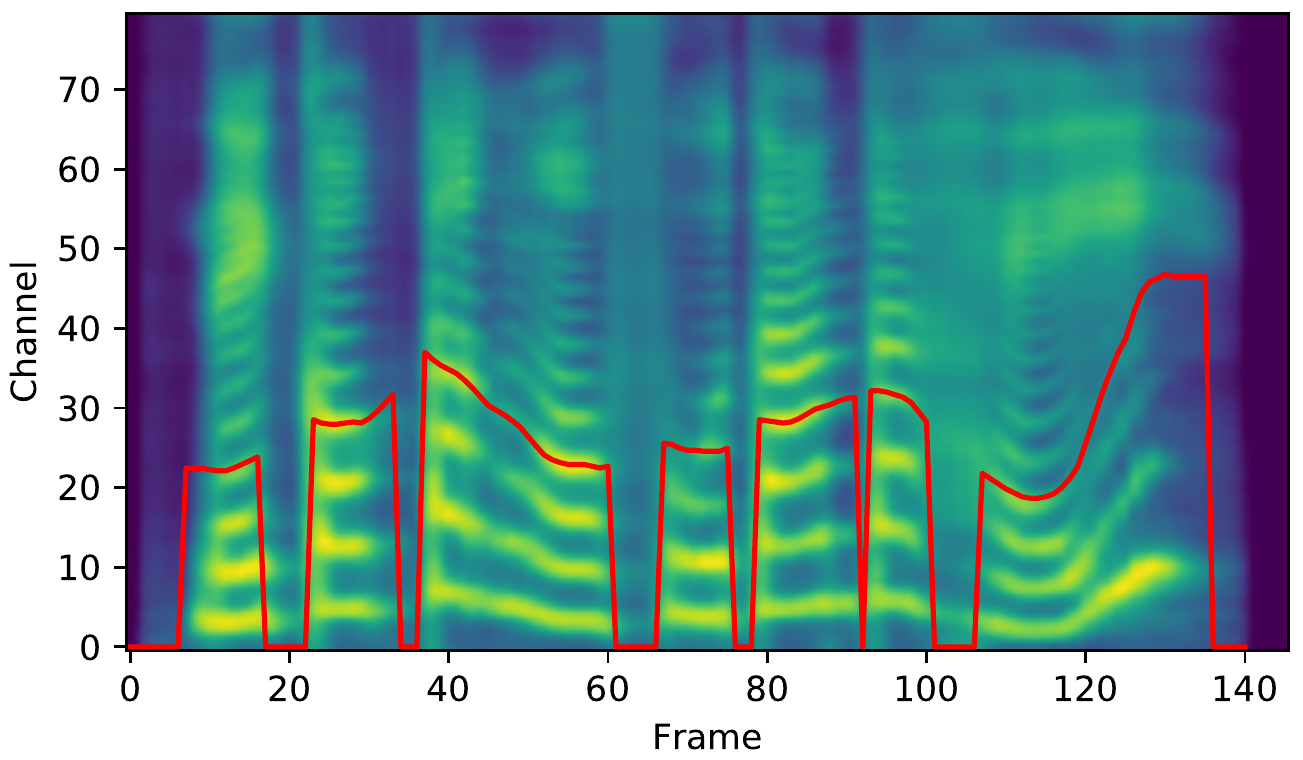}
    \caption{\normalfont ImpJoint}
    \label{fig:lmspc_impjoint}
\end{subfigure}
\begin{subfigure}{0.22\textwidth}
    \includegraphics[width=\linewidth]{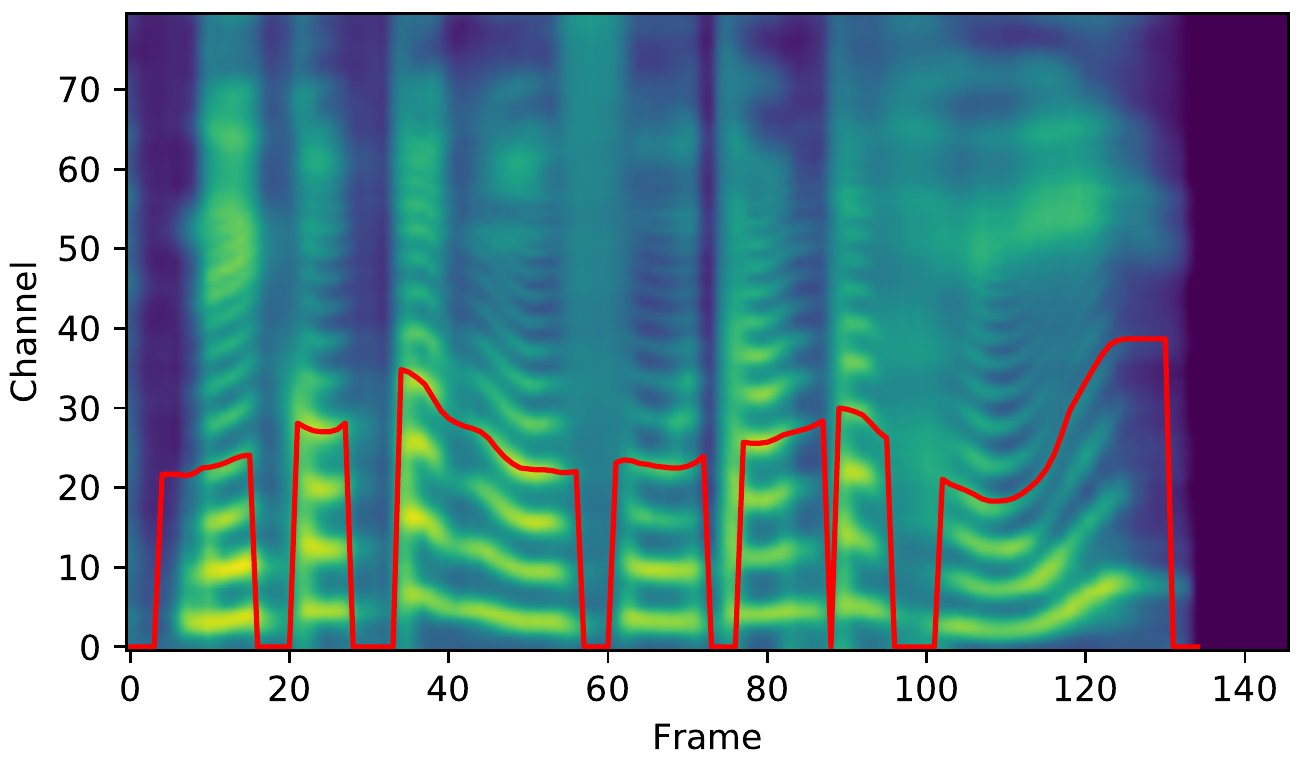}
    \caption{\normalfont ExpSep}
    \label{fig:lmspc_expsep}
\end{subfigure}
\caption{Mel-spectrograms (with F0 contours) of a declarative question. (\begin{CJK*}{UTF8}{gbsn}你觉得我负担得起？\end{CJK*}, You think I can afford it?)}
\label{fig:lmspc}
\end{figure}

A declarative question example that contains rising intonation is shown in Fig. \ref{fig:lmspc}. In the ground-truth mel-spectrogram, the pitch contour rises obviously at the end. Tacotron2 fails to render such a pattern, while other models perform much better. We also find that ExpSep can render rising intonation for long sentences. This indicates its ability in generalization.

\subsection{Objective evaluation}

Since intonation is mainly reflected by pitch contours, we apply the F0 Frame Error (FFE) \cite{chu2009reducing} for objective evaluation.
We align each synthesized mel-spectrogram and the ground truth using dynamic time wrapping (DTW) with the Mel-cepstral distortion (MCD) cost metric \cite{kubichek1993mel}. Then we synthesize waveforms, extract pitch contours, and compute the FFE. We calculate averaged FFE for each model on the test set.
The result is shown in Table \ref{tab:ffe}.

\begin{table}[ht]
\setlength\tabcolsep{4pt}
\centering
\caption{FFE of five systems on different subsets.}
\begin{tabular}{ccccc}
\toprule
System      & Sta             & Que              & DecQue           & All              \\
\midrule
Tacotron2  & 13.53\%         & 25.52\%          & 32.76\%          & 16.38\%          \\
ImpJoint     & 13.62\%         & 20.43\%          & 20.77\%          & 14.89\%          \\
ExpJoint  & \textbf{12.7\%} & 18.86\%          & 18.107\%         & \textbf{13.76\%} \\
ExpSep & 14.03\%         & 20.08\%          & 19.71\%          & 15.10\%          \\
ExpSep(GT/FT)     & 14.03\%         & \textbf{15.93\%} & \textbf{14.13\%} & 14.21\%          \\
\bottomrule
\end{tabular}
\label{tab:ffe}
\end{table}

Among the designed systems, ExpSep(GT/FT) generates declarative questions with the most similar pitch contours to the ground truth, and it performs well on normal questions. All proposed systems perform significantly better than Tacotron2 on declarative questions. We attribute this to the help of the sentence encoder/classifier.

\subsection{Subjective evaluation}

We first employ the MOS test for subjective evaluation.\footnote{Audio samples: \url{https://p1ping.github.io/RI-TTS}}
Each model is evaluated on 30 statements, 30 normal questions, and 30 declarative questions that are randomly selected from the test set. 
The raters know the sentence types of all test samples, and they need to score based on both the quality and perception of the corresponding sentence type.
For example, raters are expected to lower the score when a synthesized declarative question sounds like a statement. 
Each audio is rated by 10 native Cantonese speakers.

\begin{table}[ht]
\setlength\tabcolsep{5pt}
\centering
\caption{The MOS results of five systems on three test sets.}
\begin{tabular}{cccc}
\toprule
System     & Sta                & Que               & DecQue            \\
\midrule
Tacotron2 & 4.22$\pm$0.11          & 4.27$\pm$0.12          & 2.88$\pm$0.18          \\
ImpJoint    & 4.24$\pm$0.11          & 4.21$\pm$0.11          & 4.20$\pm$0.12          \\
ExpJoint & 4.26$\pm$0.11 & 4.23$\pm$0.12          & 4.21$\pm$0.13          \\
ExpSep & 4.21$\pm$0.11          & 4.32$\pm$0.12          & 4.33$\pm$0.12          \\
ExpSep(GT/FT)    & \textbf{4.26$\pm$0.10}          & \textbf{4.41$\pm$0.10} & \textbf{4.46$\pm$0.09} \\
\bottomrule
\end{tabular}
\label{tab:mos}
\end{table}

As Table \ref{tab:mos} shows, all designed systems with the sentence encoder/classifier have better overall performance than the baseline Tacotron2. It is because the Tacotron2 does not have rich contextual information for deciding whether to synthesize rising intonation.
The ImpJoint model, trained without sentence labels, also synthesizes most declarative questions with rising intonation. This suggests that the mel-spectrogram loss also guides the model to learn certain contextual information associated with rising intonation.

We also prepare another test set to evaluate the perception accuracy. This set contains 28 statements and 28 declarative questions. The sentence types are unknown to the raters. 
They are asked to judge whether each audio is a statement or a declarative question. We then compute the percentage of correctly perceived sentences.
The results are summarized in Table \ref{tab:acc}, which are consistent with the MOS results.

\begin{table}[ht]
\centering
\caption{Perception accuracy.}
\begin{tabular}{cccc}
\toprule
System      & Sta              & DecQue           & All              \\
\midrule
Tacotron2  & \textbf{96.43\%} & 21.83\%          & 59.13\%          \\
ImpJoint     & 94.44\%          & 85.32\%          & 89.88\%          \\
ExpJoint  & 95.63\%          & 87.70\%          & 91.67\%          \\
ExpSep & 95.63\%          & 90.87\%          & 93.25\%          \\
ExpSep(GT/FT)     & 96.03\%          & \textbf{98.41\%} & \textbf{97.22\%} \\
\bottomrule
\end{tabular}
\label{tab:acc}
\end{table}

Even though sentence labels enrich discriminative information of sentence types, ExpJoint only wins a little against ImpJoint.
This suggests that the TTS model does not make full use of the discriminative information in the joint training approach.

ExpSep has better subjective evaluation results than ExpJoint. This is because the embedding table stably catches different intonation patterns, and the TTS model can synthesize rising intonation once given the label.
Note that ExpJoint outperforms ExpSep objectively in FFE, which measures the entire pitch contour. This suggests that the sentence embedding before the classification layer also helps with the overall intonation.

As expected, ExpSep(GT) performs the best as we assume that the users can hint at the sentence type.
Fine-tuning BERT parameters proves to be effective as ExpSep(FT) predicts all testing sentences correctly.
When samples with rising intonation are insufficient to train the classifier, ExpSep can be trained with extra text data, which shows the feasibility of ExpSep.

\section{Conclusions}
\label{sec:conclusion}
\vspace{4mm}

In this paper, we propose to complement the TTS model with a BERT-based statement/question classifier.
We design different training strategies, conduct our experiments on the CanTTS Cantonese corpus and compare their performance.
All evaluated models outperform the vanilla Tacotron2 in synthesizing appropriate rising intonation.
Empirical results show that the separate training approach obtains the best performance, generalization ability, and feasibility.
In the future work, we will explore transferring rising intonation to other speakers/languages and extend the methods to model more fine-grained prosody by considering contextual information other than the intention.

\clearpage

\bibliographystyle{IEEEbib}
\bibliography{refs}

\end{document}